\documentclass[final,3p,times,12pt]{elsarticle}

\usepackage{graphicx,graphics}
\usepackage{multirow,array,bm,enumerate}
\usepackage{amssymb,epsfig,amsmath,comment,hyperref,verbatim}
\usepackage{subfigure,setspace}
\usepackage{textcomp}
\usepackage{url}
\usepackage{color}

\usepackage{amsmath,graphicx}
\usepackage{graphics,comment}
\usepackage{tikz}
\usepackage{tikz}
\usetikzlibrary{arrows,patterns}
\usetikzlibrary{positioning}

\usepackage{lineno,hyperref}
\usepackage{amsmath,amssymb,amsfonts}
\usepackage{algorithmic}
\usepackage{textcomp}

\def\BibTeX{{\rm B\kern-.05em{\sc i\kern-.025em b}\kern-.08em
    T\kern-.1667em\lower.7ex\hbox{E}\kern-.125emX}}

\usepackage[calc,showdow,english]{datetime2}
\usepackage{tabularx}
\usepackage[utf8]{inputenc}
\usepackage{rotating}

\usepackage{booktabs}
\usepackage{multirow}

\begin{document}

\begin{frontmatter}

\title{Intelligibility of Speech in Noise: Investigating Contribution of Magnitude and Phase Spectra}

\author{Bhanu Teja Nellore$^1$, Sudarsana Reddy Kadiri$^2$, Rohit Kumar$^3$, Karan Nathwani
$^4$, and Suryakanth V Gangashetty$^5$}
\address{$^1$Jio AICoE, Hyderabad, India\\$^2$Signal Analysis and Interpretation Laboratory, University of Southern California, Los Angeles, USA\\$^3$National Institute of Technology, Patna, India\\ $^4$Indian Institute of Technology, Jammu, India\\ $^5$Koneru Lakshmaiah Education Foundation, Vaddeswaram, Guntur District, Andhra Pradesh, India}

\begin{abstract}
It is well known that intelligibility of speech reduces in the presence of ambient noise. However, studies show that all sounds are not affected uniformly (or equally) and that vowels are more robust to noise than consonants. In this study, intelligibility of various consonants is assessed and analyzed in stationary white noise and non-stationary babble noise conditions. Specifically, this study investigates the individual contribution of magnitude and phase spectra of a given speech signal on human speech recognition of consonants in noisy conditions. In this regard, three experiments are carried out. In experiment 1, clean signal, signal reconstructed with only magnitude spectrum information (magnitude only signal) and signal reconstructed with only phase spectrum information (phase only signal) are assessed for intelligibility. In experiment 2, noise is added to clean speech. From noisy speech, phase only signal and magnitude only signal are reconstructed and intelligibility tests are performed for all these three signals. In experiment 3, noise is added directly to the magnitude only and phase only signals reconstructed from clean speech and their intelligibility is assessed. Results of these experiments show that magnitude spectrum contributes more to intelligibility in clean condition than phase spectrum, while information from phase spectrum is more robust in noisy conditions. It is also observed that, among consonants, nasals are more susceptible to noise whereas fricatives and approximants were observed to be comparatively more robust.
\end{abstract}


\end{frontmatter}


\section{Introduction}
\label{sec:intro}
Speech signal processing using phase information finds application in many areas such as speech analysis \cite{IF_suda,suda_Thesis}, speech enhancement \cite{gerkmann2015phase, mowlaee2017iterative, gaich2015on}, locating burst onsets \cite{bhanu2017locating}, speech coding \cite{agiomyrgiannakis2009wrapped, pobloth2004perceptual}, digital speech watermarking \cite{hernaez2014speech} and automatic speech/speaker recognition \cite{mccowan2011the, vijayan2014epoch}. In automatic speech recognition (ASR) systems, traditionally magnitude spectrum is used to extract feature representations. Some studies suggest phase information can be used to build ASR systems, even with smaller short time Fourier transform (STFT) window duration \cite{alsteris2007short,alsteris2004importance}. More details about the phase processing in speech can be found in \cite{alsteris2007short}. One of the goals of this paper is to study and compare the contribution of phase spectrum with that of magnitude spectrum w.r.t. human speech recognition. 

Speech sounds can be broadly classified into two categories, namely, vowels and consonants. Studies suggest that vowels are more intelligible than consonants in noisy conditions \cite{allen2007consonant} and hence conversely, consonants are more susceptible to noise than vowels due to reasons such as low energy, noise-like characteristics in stops, fricatives, etc. Since speech intelligibility in noisy conditions is highly dependent on consonant intelligibility, it is important that we study the intelligibility of consonants in noise. The goals of the current study can be outlined as follows:
\begin{itemize}
\item Study the contribution of different classes of consonants to speech intelligibility in noisy conditions.
\item Under similar conditions, also study the individual contributions of magnitude and phase spectra on speech intelligibility.
\end{itemize}

\begin{figure*}
\centering
\includegraphics[width=15cm,height=10cm]{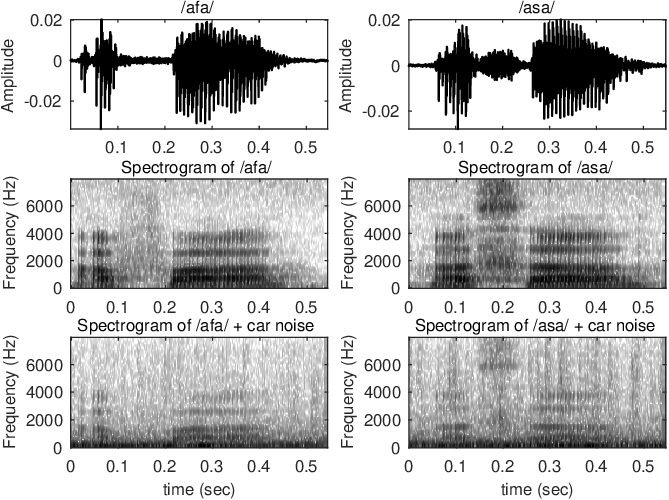}
\caption{Interaction of /f/ and /s/ respectively with car noise at -10 dB SNR.}
\label{s_f_spectrograms}
\end{figure*}

It has been shown in \cite{allen2007consonant, allen2008consonant, allen2014across} that different speech sounds are affected differently in a given noise condition. Let's consider an example presented in Fig. \ref{s_f_spectrograms}, where it illustrates the interaction of two fricatives, namely, /f/ and /s/ with car noise (-10 dB SNR). /f/ is a labio-dental fricative while /s/ is an alveloar fricative. From Fig. \ref{s_f_spectrograms} the following observations can be made. /f/ has relatively low acoustic energy compared to /s/ and the energy distributions  of /f/ and /s/ in spectral domain are different from each other. When these consonants interact with car noise, the spectral information in /s/ is more robust compared to /f/. As a result /s/ is more intelligible in car noise environment. Therefore it is intuitive to design speech modification algorithms by giving attention to  individual sound units in a speech signal for improving intelligibility. 

Studies on contribution of consonants in clean conditions have been done in \cite{paliwal2005on, li2008contribution, nivedita2018significance}. In this study, we are investigating the perceptual robustness of various consonants in two noise conditions, namely, white (stationary) noise and babble (non-stationary) noise. For each consonant three different signals are considered corresponding to three components mentioned above:

\begin{itemize}
\item clean original signal,
\item signal reconstructed using information of magnitude spectrum preserved and phase spectrum destroyed (hereby called magnitude only signal), and
\item signal reconstructed using information of magnitude spectrum destroyed and phase spectrum preserved (hereby called phase only signal)
\end{itemize}

Three experiments are conducted:
	\begin{itemize}
		\item In Experiment 1, intelligibility of clean signal, magnitude only signal and phase only signal are evaluated for all consonants.
		\item In Experiment 2, noisy speech signal in constructed using clean original signal and noise signal. From this noisy speech signal, magnitude only and phase only signals are derived. Intelligibility of these signals is evaluated for all consonants.
		\item In Experiment 3, noise is added to magnitude only and phase only signals considerred in Experiment 1. Intelligibility of these signals is evaluated for all consonants.
	\end{itemize}

In all the above experiments, intelligibility performance of consonants is analysed w.r.to overall consonant identification rate and identification rate based Manner of Articulation (MoA) of consonants (such as stops\&affricates, fricatives, liquides\&glides and nasals).

Therefore, this study analyses:
\begin{itemize}
\item perceptual robustness of various consonants in specified noise environments and 
\item under similar conditions
\begin{itemize}
\item perceptual robustness of magnitude spectrum contribution of various consonants and
\item perceptual robustness of phase spectrum contribution of various consonants.
\end{itemize}
\end{itemize}

This paper is organized as follows. Section~\ref{sec:methods} describes STFT analysis, modification and synthesis (AMS) method for extracting magnitude only and phase only signals. Section~\ref{sec:exp} elaborates details of experiments 1, 2 and 3 as mentioned above. Section~\ref{sec:results} discusses the results of subjective perception. Section~\ref{sec:discussion} discusses the results of objective metrics. Summary and conclusion of the study is presented in Section~\ref{sec:summary}.

\section{Short Time Fourier Transform based Analysis, Modification and Synthesis}
\label{sec:methods}

Even though speech is a non-stationary signal, it can be considered to be quasi-stationary within a short time frame (20-40 ms) and therefore can be processed through STFT \cite{quatieri2006discrete}. The STFT of a speech signal $s[n]$ is given by: 
\begin{equation} \label{STFT}
S(n,k) = \sum_{m=-\infty}^{\infty}s[m]w[n-m]e^{-j\frac{2\pi}{N}k n} ,
\end{equation}
where $w[n]$ is window function.

where 
\begin{itemize}
\item[ ] $S(n,k)$ is spectro temporal representation of $s[n]$,
\item[ ] $n$, $k$ are frame and frequency indices respectively,
\item[ ] $w[n]$ is window,
\item[ ] $m$ represents time shift in window, and
\item[ ] N is number of points in Discrete Fourier Transform (DFT).
\end{itemize}

$S(n,k)$ can be decomposed as follows
\begin{equation}\label{magph}
S(n,k) = \left | S(n,k) \right |e^{j\phi(n,k)} ,
\end{equation}
where $\left | S(n,k) \right |$ is STFT magnitude spectrum and $\phi(n,k)$ is STFT phase spectrum.

The signal s[n] can be completely characterized/reconstructed using its magnitude and phase spectra. The aim of this study is to determine the contribution of magnitude spectrum and phase spectrum for intelligibility of consonant sounds in clean and noisy conditions. For this purpose speech stimuli are created by preserving information either in magnitude or phase spectrum and destroying the information in the other. This method was first proposed in \cite{liu1997effects} and is termed as analysis, modification and sysnthesis technique (AMS).

\subsection{Phase only signal}
In order to reconstruct speech with only phase spectrum information, the signal is processed through STFT analysis using Eq. (\ref{STFT}), decomposed as shown in Eq. (\ref{magph}). The phase spectrum is retained while magnitude spectrum is made unity, i.e.,

\begin{equation}\label{phaseonlyeq}
S_{ph}(n,k) = 1e^{j\phi(n,k)} .
\end{equation}

From $S_{ph}(n,k)$, the signal $s_{ph}[n]$ is reconstructed using STFT synthesis (overlap add method) equation given below.

\begin{equation}
s_{ph}[n] = \frac{1}{W[0]}\sum_{p=-\infty}^{\infty}\left [ \sum_{k=0}^{N-1}S_{ph}(p,k)e^{j\frac{2\pi}{n}kn} \right ] .
\end{equation}

From here on $s_{ph}[n]$ is referred as phase only signal.

\subsection{Magnitude only signal} 
In order to reconstruct speech with only magnitude spectrum information, the signal is processed through STFT analysis using Eq. (\ref{STFT}), decomposed as shown in Eq. (\ref{magph}). The magnitude spectrum is retained while phase spectrum is randomized (random numbers from uniform distribution ranging from 0 to $2\pi$), $\phi_{random}(n,k)$.
\begin{equation}
S_{mag}(n,k) = \left | S(n,k) \right |e^{j\phi_{random}(n,k)} .
\end{equation}

From $S_{mag}(n,k)$, the signal $s_{mag}[n]$ is reconstructed using STFT synthesis (overlap add method) equation given below.

\begin{equation}
s_{mag}[n] = \frac{1}{W[0]}\sum_{p=-\infty}^{\infty}\left [ \sum_{k=0}^{N-1}S_{mag}(p,k)e^{j\frac{2\pi}{n}kn} \right ] .
\end{equation}

From here on $s_{mag}[n]$ is referred as magnitude only signal. 

Illustration of clean signal ($s[n]$), magnitude only signal ($s_{mag}[n]$) and phase only signal ($s_{ph}[n]$) is shown in Figure 2. It is to be noted, for intelligible reconstruction of magnitude only and phase only signals, choice of parameters such as window shape, window duration, window shift and number of points in DFT are crucial \cite{alsteris2004importance,paliwal2008effect,paliwal2003usefulness,lecumberri2008non}. According to our observation, clean signal is most intelligible. Although magnitude only and phase only signals sound distorted compared to clean signal, they are also intelligible as well.

\begin{figure}[t] 
\label{Illustration}
\centering
\includegraphics[width =12cm,height=10cm]{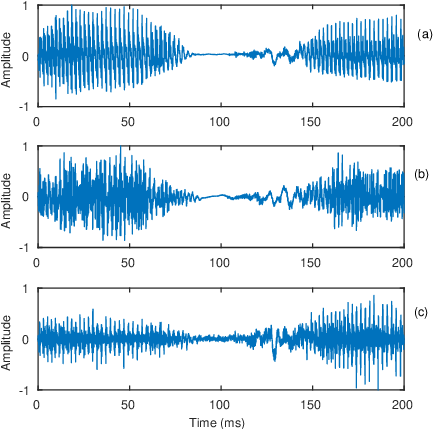}
\caption{Illustration of (a) clean, (b) magnitude only and (c) phase only signals for a portion of the utterance /aka/.}
\vspace{-0.4cm}
\end{figure}

\begin{table}[htb] \label{STFTSpecs}
\centering
\caption{Analysis and synthesis window parameters used to reconstruct magnitude only and phase only signals in clean speech. Here N is the number of DFT points.} 
\begin{tabular}{|l||l|l|l|l|}
\hline
                                                         & Window Type & Size ($T_w$)  & Shift & Padding \\ \hline \hline
\begin{tabular}[c]{@{}l@{}}Magnitude\\ Only\end{tabular} & Hamming     & 32ms  & $T_w/8$  & 2N      \\ \hline
\begin{tabular}[c]{@{}l@{}}Phase\\ Only\end{tabular}     & Rectangular & 512ms & $T_w/8$  & 2N      \\ \hline
\end{tabular}
\end{table}

\subsection{AMS design parameters}
The intelligibility of $s_{mag}[n]$ and $s_{ph}[n]$ derived using AMS method critically depends on choice of certain design parameters such as:
\begin{itemize}
\item[-] window type: the choice of window function in Eq. \ref{STFT},
\item[-] window size: duration of window function used in Eq. \ref{STFT},
\item[-] window shift: temporal shift in duration of the window function. This parameter decides the sampling rate of STFT and should be chosen such that anti-aliasing affect does not take place.
\item[-] padding: zero padding is essential to avoid aliasing effects. In all the signals generated in this study, while the original frame size is N, N more zeros are appended before applying DFT.
\end{itemize}

\begin{figure} 
\centering
\includegraphics[width =10cm,height=7cm]{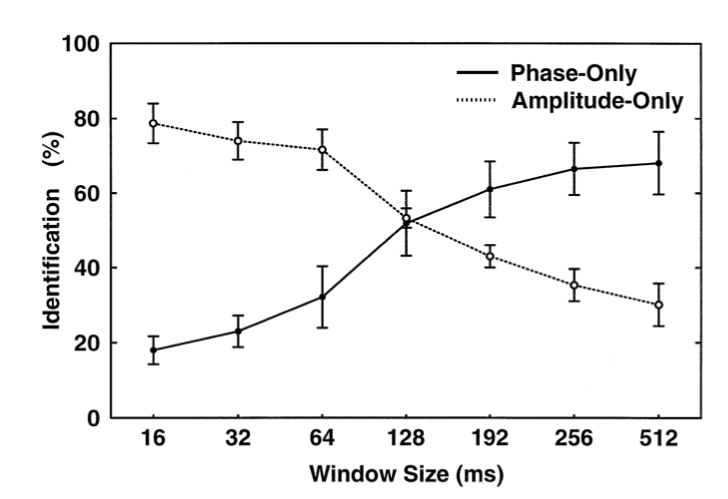}
\caption{Identification rate as a function of window size for magnitude only and phase only signals (adapted from \cite{liu1997effects}).}
\label{phase_perception}
\end{figure}

The values of above parameters need to be chosen such that the synthesized signals $s_{mag}[n]$ and $s_{ph}[n]$ are intelligible. It is well known that magnitude only signals are most intelligible for short window durations and phase only signals are most intelligible for long window durations. In studies \cite{liu1997effects} and \cite{paliwal2005on}, intelligibility tests are performed on magnitude only and phase only signals with varying window lengths. In \cite{liu1997effects}, authors used 6 stop consonants ([p,t,k,b,d,g]) for the study and the results obtained by varying the window size is shown in Fig. \ref{phase_perception}. In \cite{paliwal2005on}, authors used 16 consonants, and the AMS design parameters that produced best intelligibility scores on magnitude only and phase only signals are presented in Table 1. This configuration is used in the current study to generate $s_{mag}[n]$ and $s_{ph}[n]$. That is, to generate $s_{mag}[n]$ a hamming window of 32 ms duration with a shift of 4 ms is used to synthesize $s_{mag}[n]$. Similarly, a rectangular window of 512 ms duration with a shift of 64 ms is used to synthesize $s_{ph}[n]$. In both cases, a set of N zeros are padded to windowed frame while calculating N point DFT.

\begin{figure}[h!]
\centering
\includegraphics[width = 10cm]{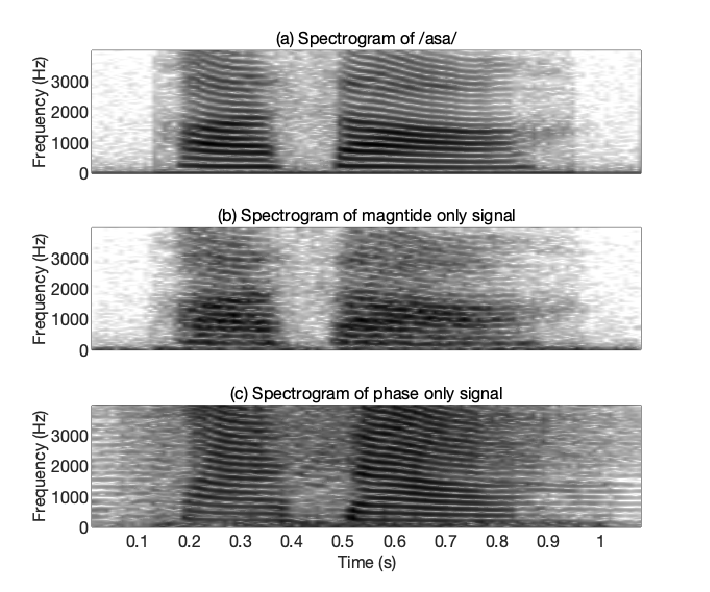}
\caption{Spectrograms of clean, magnitude only and phase only signals.}
\label{all spectrograms}
\end{figure}
	
Fig. \ref{all spectrograms} shows spectrogram of clean signal of utterance /asa/, followed by spectrograms of magnitude only and phase only signal spectrograms derived from /asa/. Sound related information can still be noticed in magnitude only and phase only spectrograms.

\section{Experimental Details and Results}
\label{sec:exp}

\subsection{Stimuli}
Listeners are presented with IEEE Vowel-Consonant-Vowel (VCV) utterances \cite{loizou2007speech} in a randomized fashion. While the vowel in all these utterances remains same (/a/), consonants change. The list of consonants that are used in this study is given in Table 2. In total there are 19 consonants. By using VCV utterances instead of words, it is not possible to guess consonants based on context.

\subsection{Experimental Protocol}
20 listeners (10 male and 10 female) belonging to L2-normal hearing category aged between 20 to 30 years participated in the listening experiments. All of them belong to different native backgrounds from to different parts of India and all of them are students of IIIT Hyderabad.
Before beginning the listening tests, each participant is trained to map graphical symbols presented in Table 2 to their respective /aCa/ utterance. After the participant declares that he/she is confident of performing this mapping without assistance the listening tests begin. The listening tests were conducted in a normal lab environment by  using Sennheiser over-the-ear headphones.

While performing experiments, all VCV tokens mentioned in Table 2 were presented to each listener in a randomized fashion. A listener was allowed to take as many breaks as required by him/her while performing listening tests. Also, the participants are allowed to listen to the utterance as many times as he/she wishes before making decision. On an average each listener took more than 2 hours to complete the experiments.

A listener can make one of the following three possible entries while presented with a VCV token: He/she may predict the right consonant, he/she may predict a wrong consonant or he/she may declare that the consonant is not intelligible.

\subsection{Experiments}
Intelligibility of consonants in clean and noisy conditions, and also individual contributions of magnitude and phase spectra in this regard are evaluated by performing three different listening experiments. The details of these experiments are explained below.

For deriving magnitude only and phase only signals, STFT parameters are adopted from \cite{paliwal2005on}. These details are given in Table 1. For simulating noisy speech, noise files are added from NOISEX database \cite{NOISEX}. In this study, effects of stationary noise (white) and non-stationary noise (babble) on consonant intelligibility are assessed.  For all the experiments, mean consonant identification rate is presented for clean signal, magnitude only signal, and phase only signal in white and babble noise environments across 0 dB and -5 dB SNRs, respectively.

\begin{table}[htb] \label{consonantlist}
\centering
\caption{List of consonants used in this study} 
\begin{tabular}{|l||l|}
\hline
Category                                                           & Consonants                                                             \\ \hline \hline
\begin{tabular}[c]{@{}l@{}}Stops and \\  Affricates\end{tabular} & \begin{tabular}[c]{@{}l@{}}p, t, k, b, d, g,\\ dh, dj, ch\end{tabular} \\ \hline
Fricatives                                                         & s, sh, f, z                                                            \\ \hline
\begin{tabular}[c]{@{}l@{}}Liquids and \\ Glides\end{tabular}   & l, r, v, y                                                             \\ \hline
Nasals                                                             & m, n                                                                   \\ \hline
\end{tabular}
\end{table}

\subsubsection{Experiment 1}
The purpose of this experiment is to test contribution of magnitude and phase components of a signal in clean environment.
Along with clean speech signal, magnitude only signal and phase only signal are considered. These 3 signals are presented to listeners in random order. Figure 5 shows identification rates obtained in this experiment. 

For all the experiments in this study, we present the results in the following manner. Panel (a) presents identification rates of clean speech, magnitude only signal, and phase only signal respectively. The remaining panels present breakup of identification rates for above signals w.r.to sound categories mentioned in Table 2. The color coding used in all figures is mentioned in Table 3.

\begin{table}[]
\centering
\caption{Color coding scheme for Figures 5-9.} 
\vspace{0.1cm}
\begin{tabular}{|l||l|l|}
\hline 
Panel                             & Color      & Signal                \\ \hline \hline
\multirow{3}{*}{Panel (a)}        & Dark Blue  & Clean signal          \\ \cline{2-3} 
                                  & Light Blue & Magnitude only signal \\ \cline{2-3} 
                                  & Yellow     & Phase only signal     \\ \hline
\multirow{4}{*}{Remaining Panels} & DarkBlue   & Stops and Affricates  \\ \cline{2-3} 
                                  & Light Blue & Fricatives            \\ \cline{2-3} 
                                  & Green      & Liquids and Glides    \\ \cline{2-3} 
                                  & Yellow     & Nasals                \\ \hline
\end{tabular}
\end{table}

\begin{figure*}[ht]
\centering
\includegraphics[width=14cm,height=10cm]{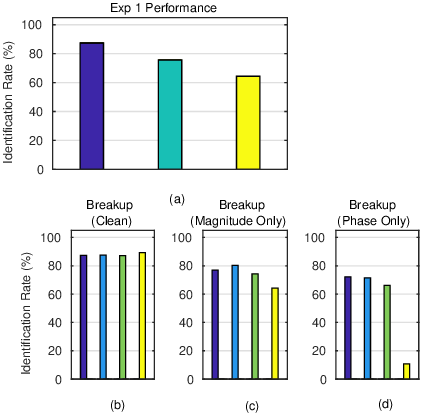}
\caption{Speech intelligibility performance for experiment 1. (a) Identification rates of clean, magnitude only and phase only signals. (b, c, d) Breakup of their identification rates according to consonant categories mentioned in Table 2.}
\vspace{-0.1cm}
\end{figure*}

\begin{figure*}
\centering
\begin{minipage}{\linewidth}
\includegraphics[width=14cm,height=10cm]{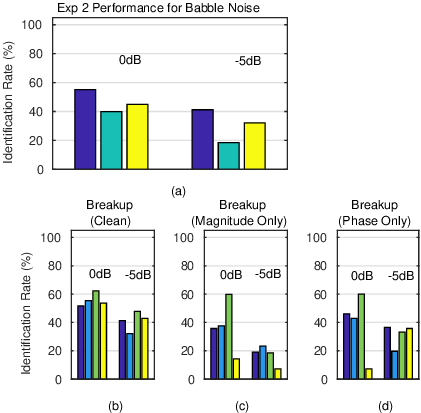}
\caption{Speech intelligibility performance for experiment 2 in babble noise condition at 0 dB and -5 dB SNR levels. (a) Identification rates of clean, magnitude only and phase only signals. (b, c, d) Breakup of their identification rates according to consonant categories mentioned in Table 2.}
\end{minipage}

\begin{minipage}{\linewidth}
\vspace{0.3cm}
\includegraphics[width=14cm,height=10cm]{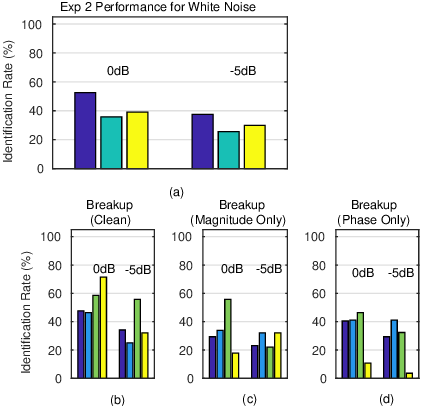}
\caption{Speech intelligibility performance for experiment 2 in white noise condition 0 dB and -5 dB SNR levels. (a) Identification rates of clean, magnitude only and phase only signals. (b, c, d) Breakup of their identification rates according to consonant categories mentioned in Table 2.}
\end{minipage}
\end{figure*}

\begin{figure*}
\centering 
\begin{minipage}{\linewidth}
\includegraphics[width=14cm,height=10cm]{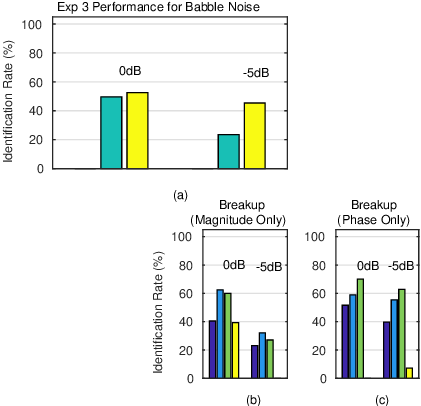}
\caption{Speech intelligibility performance for experiment 3 in babble noise condition 0 dB and -5 dB SNR levels. (a) Identification rates of magnitude only and phase only signals. (b, c) Breakup of their identification rates according to consonant categories mentioned in Table 2.\newline}
\end{minipage}

\begin{minipage}{\linewidth}
\includegraphics[width=14cm,height=10cm]{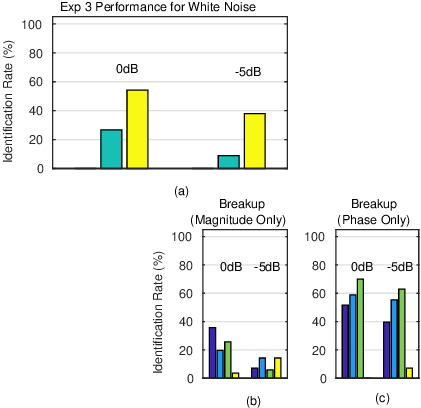}
\caption{Speech intelligibility performance for experiment 3 in white noise condition 0 dB and -5 dB SNR levels. (a) Identification rates of magnitude only and phase only signals. (b, c) Breakup of their identification rates according to consonant categories mentioned in Table 2. \newline}
\end{minipage}
\vspace{-0.3cm}
\end{figure*}

\subsubsection{Experiment 2}
This experiment assesses intelligibility of speech and contribution of magnitude and phase spectra in noisy conditions. At first, clean speech is corrupted with noise. Magnitude only and phase only signals are derived from this noise corrupted speech. As mentioned earlier, two types of noises are assessed separately, viz., white and babble at 0 dB and -5 dB SNRs, respectively. The intelligibility performance of experiment 2 is presented in Figures 6 and 7. 

\subsubsection{Experiment 3}
The purpose of this experiment is to test the robustness of magnitude only and phase only signals. $s_{mag}[n]$ and $s_{ph}[n]$ are derived from clean speech and then corrupted with two types of noises (white and babble at 0 dB and -5 dB SNRs). Results of this experiment are shown in Figures 8 and 9. It can be noted that clean speech is not assessed in this experiment and hence their respective identification rates are not present in these figures.


For Experiment 1, each listener was presented with $19\times 3=57$ VCV tokens. For Experiment 2, each listener was presented with $19\times 6=114$ VCV tokens. For Experiment 3, each listener was presented with $19\times 4=76$ VCV tokens. Therefore, in total each listener was presented with 247 VCV tokens. 
\section{Results and Discussion of subjective perception experiments}
\label{sec:results}

Figure 5(a) shows identification rate for clean, magnitude only, and phase only signals (experiment 1).  It can be observed that the performance for clean case is  not 100\%. Experiment 1 is done only on clean signal (without noise). According to our analysis, errors in intelligibility arise mainly due to confusions between consonants such as /dj/ and /ch/; /z/, /r/ and /l/; and also sometimes between /m/ and /n/. Since these confusions did not arise due to noise, they can be attributed to factors such as accent diversity. It can be noted here that IEEE VCV dataset is recorded using English speaker and listening tests are done by non-English (Indian) listeners. According to the results of experiment 1, clean signal is most intelligible followed by magnitude only signal, and then followed by phase only signal. According to sound categories, nasals are most affected intelligibility-wise due to signal reconstruction (Figures 5 (b), (c) and (d)).

Although magnitude only signal is more intelligible in clean condition (Figure 5(a)) than phase only signal, phase only signal is more robust than magnitude only signal in noisy conditions (Figures 6-9 (a)), especially so at lower SNR levels (-5 dB). When the reconstructed signal is corrupted with noise (Experiment 3), phase only signal is categorically more intelligible than magnitude only signal for both babble and white noise conditions (Figures 8(a) and 9(a)). In noisy conditions (experiments 2 and 3), nasals are most affected consonants while fricatives and approximants are relatively more robust (Figures 6,7 (c,d) and Figures 8,9 (b,c) respectively).

When clean speech signal is corrupted by noise and $s_{mag}[n]$ and $s_{ph}[n]$ are derived from it (experiment 2), the intelligibility of phase only signal is only slightly better than it's magnitude only counterpart (Figures 6(a) and 7(a)). The overall intelligibility performance across all experiments is affected more by white noise than by babble noise (Figures 6-9 (a)).

\begin{table*}
\centering
\caption{Objective Scores of Clean signal, Magnitude Only Signal and Phase Only Signal under different noise conditions.}
\vspace{0.4cm}
\label{Obj_metrics}
\resizebox{12cm}{4cm}{
\begin{tabular}{lllclll}
\hline
\multicolumn{2}{|l|}{Noise}                         & \multicolumn{1}{l|}{Type of signal}        & \multicolumn{1}{l|}{Subjective Results} & \multicolumn{1}{l|}{CSII} & \multicolumn{1}{l|}{Ext-SII} & \multicolumn{1}{l|}{GP}    \\ \hline \hline
\multicolumn{2}{|l|}{\multirow{3}{*}{White 0 dB}}   & \multicolumn{1}{l|}{Signal}                & \multicolumn{1}{c|}{52.5}               & \multicolumn{1}{l|}{0.63} & \multicolumn{1}{l|}{0.32}    & \multicolumn{1}{l|}{16.93} \\ \cline{3-7} 
\multicolumn{2}{|l|}{}                              & \multicolumn{1}{l|}{Magnitude only signal} & \multicolumn{1}{c|}{26.79}              & \multicolumn{1}{l|}{0.63} & \multicolumn{1}{l|}{0.33}    & \multicolumn{1}{l|}{17.62} \\ \cline{3-7} 
\multicolumn{2}{|l|}{}                              & \multicolumn{1}{l|}{Phase only signal}     & \multicolumn{1}{c|}{54.29}              & \multicolumn{1}{l|}{0.63} & \multicolumn{1}{l|}{0.47}    & \multicolumn{1}{l|}{17.71} \\ \hline
                               &                    &                                            &                                         &                           &                              &                            \\ \hline
\multicolumn{2}{|l|}{\multirow{3}{*}{White -5 dB}}  & \multicolumn{1}{l|}{Signal}                & \multicolumn{1}{c|}{37.5}               & \multicolumn{1}{l|}{0.48} & \multicolumn{1}{l|}{0.26}    & \multicolumn{1}{l|}{11.01} \\ \cline{3-7} 
\multicolumn{2}{|l|}{}                              & \multicolumn{1}{l|}{Magnitude only signal} & \multicolumn{1}{c|}{8.98}               & \multicolumn{1}{l|}{0.48} & \multicolumn{1}{l|}{0.26}    & \multicolumn{1}{l|}{10.74} \\ \cline{3-7} 
\multicolumn{2}{|l|}{}                              & \multicolumn{1}{l|}{Phase only signal}     & \multicolumn{1}{c|}{37.99}              & \multicolumn{1}{l|}{0.47} & \multicolumn{1}{l|}{0.36}    & \multicolumn{1}{l|}{4.63}  \\ \hline
                    &                    &                                            &                                         &                           &                              &                            \\ \hline
\multicolumn{2}{|l|}{\multirow{3}{*}{Babble 0 dB}}  & \multicolumn{1}{l|}{Signal}                & \multicolumn{1}{c|}{55.2}               & \multicolumn{1}{l|}{0.59} & \multicolumn{1}{l|}{0.37}    & \multicolumn{1}{l|}{16.96} \\ \cline{3-7} 
\multicolumn{2}{|l|}{}                              & \multicolumn{1}{l|}{Magnitude only signal} & \multicolumn{1}{c|}{49.64}              & \multicolumn{1}{l|}{0.58} & \multicolumn{1}{l|}{0.38}    & \multicolumn{1}{l|}{16.32} \\ \cline{3-7} 
\multicolumn{2}{|l|}{}                              & \multicolumn{1}{l|}{Phase only signal}     & \multicolumn{1}{c|}{52.5}               & \multicolumn{1}{l|}{0.6}  & \multicolumn{1}{l|}{0.54}    & \multicolumn{1}{l|}{34.91} \\ \hline
                      &                    &                                            &                                         &                           &                              &                            \\ \hline
\multicolumn{2}{|l|}{\multirow{3}{*}{Babble -5 dB}} & \multicolumn{1}{l|}{Signal}                & \multicolumn{1}{c|}{41.22}              & \multicolumn{1}{l|}{0.45} & \multicolumn{1}{l|}{0.3}     & \multicolumn{1}{l|}{10.26} \\ \cline{3-7} 
\multicolumn{2}{|l|}{}                              & \multicolumn{1}{l|}{Magnitude only signal} & \multicolumn{1}{c|}{23.51}              & \multicolumn{1}{l|}{0.45} & \multicolumn{1}{l|}{0.3}     & \multicolumn{1}{l|}{9.73}  \\ \cline{3-7} 
\multicolumn{2}{|l|}{}                              & \multicolumn{1}{l|}{Phase only signal}     & \multicolumn{1}{c|}{45.36}              & \multicolumn{1}{l|}{0.45} & \multicolumn{1}{l|}{0.45}    & \multicolumn{1}{l|}{26.66} \\ \hline
\end{tabular}}
\end{table*}

\section{Results and Discussion with Objective Measures}
\label{sec:discussion}

From subjective measures, it was found that phase only reconstructed signals was shown to preserve better speech intelligibility in noisy conditions compared to original speech signal and magnitude only reconstructed signals. Apart from subjective measures, the objective measures of intelligibility like glimpse proportion (GP) \cite{cooke2006glimpsing}, extended speech intelligibility index (Ext-SII) \cite{rhebergen2005speech} and coherence speech intelligibility index (CSII) \cite{kates2005coherence} for the noisy signals are also measured and the results are given in Table~\ref{Obj_metrics}. From the table, it can be observed that for both types of noises, Ext-SII and GP show higher scores for phase only reconstructed signals when compared to original signals and magnitude only reconstructed signals, which comply to the subjective results. Hence, it appears that Ext-SII and GP measures are relevant to evaluating the intelligibility of phase only reconstructed signals. On the other hand, the scores of CSII seems to be nearly in the same range or equal for all the three signals (original, magnitude only signals, and phase only signals) even though there exists variations in subjective measures. According to the results of CSII, it appears that this measure may not be relevant for evaluating the intelligibility of magnitude only reconstructed signals and phase only reconstructed signals. However, this needs proper investigation, which is beyond the scope of the present study.

\section{Summary and Conclusion}
\label{sec:summary}

In this study human speech recognition tests were performed to assess the individual contributions of magnitude and phase spectra on consonant intelligibility in noisy conditions. According to the results, magnitude only signal is more intelligible than phase only signal in clean condition. In noisy conditions, phase only signal is more robust than magnitude only signal. Results also suggest that consonant intelligibility is affected more by stationary white noise than by non-stationary babble noise. Among consonants, nasals are more susceptible to noise whereas fricatives and approximants were observed to be comparatively more robust. As a part of further study, we would like to use the above knowledge to design speech enhancement methods and produce noise robust intelligible speech signals. 

\bibliography{bhanu_master.bib}

\end{document}